\documentclass[12pt, draftclsnofoot,onecolumn]{IEEEtran}
\IEEEoverridecommandlockouts
\usepackage[english]{babel}
\usepackage{blindtext}
\usepackage{algorithm}
\usepackage{algorithmic}
\usepackage{graphicx}
\usepackage{subfigure}
\usepackage{amsmath}
\usepackage{amssymb}
\usepackage{cases}
\usepackage{setspace}
\usepackage{multirow}
\usepackage{color}

\begin{document}

\title{Deep Reinforcement Learning for Multi-Agent Power Control in Heterogeneous Networks}
\author{\authorblockN{Lin Zhang and Ying-Chang Liang, \IEEEmembership{Fellow, IEEE}}


\thanks{L. Zhang is with the Key Laboratory on Communications, University of Electronic Science and Technology of China (UESTC), Chengdu, China (emails: linzhang1913@gmail.com).}
\thanks{Y.-C. Liang is with the Center for Intelligent Networking and Communications (CINC), University of Electronic Science and Technology of China (UESTC), Chengdu, China (email: liangyc@ieee.org).}
%

}

  \maketitle



\begin{abstract}

We consider a typical heterogeneous network (HetNet), in which multiple access points (APs) are deployed to serve users by reusing the same spectrum band. Since different APs and users may cause severe interference to each other, advanced power control techniques are needed to manage the interference and enhance the sum-rate of the whole network. Conventional power control techniques first collect instantaneous global channel state information (CSI) and then calculate sub-optimal solutions. Nevertheless, it is challenging to collect instantaneous global CSI in the HetNet, in which global CSI typically changes fast. In this paper, we exploit deep reinforcement learning (DRL) to design a multi-agent power control algorithm in the HetNet. To be specific, by treating each AP as an agent with a local deep neural network (DNN), we propose a multiple-actor-shared-critic (MASC) method to train the local DNNs separately in an online trial-and-error manner. With the proposed algorithm, each AP can independently use the local DNN to control the transmit power with only local observations. Simulations results show that the proposed algorithm outperforms the conventional power control algorithms in terms of both the converged average sum-rate and the computational complexity.


\end{abstract}

\newpage

%
\section{Introduction}

Driven by ubiquitous wireless-devices like smart phones and tablets, wireless data traffics have been dramatically increasing in recent years \cite{6GBe}-\cite{Lin_SS_2017}. These wireless traffics lay heavy burden on conventional cellular networks, in which a macro \emph{base station} (BS) is deployed to provide wireless access services for all the users within the macro-cell. As an alternative, the \emph{heterogeneous network} (HetNet) is proposed by planning small cells in the macro-cell. Typical small cells include Pico-cell and Femto-cell, which are able to provide flexible wireless access services for users with heterogeneous demands. It has been demonstrated that small cells can effectively offload the wireless traffics of the macro BS, provided that the small cells are well coordinated \cite{HetNet_survey1}.

Due to the spectrum scarcity issue, it is inefficient to assign orthogonal spectrum resources to all the cells (including macro-cell and small cells). Then, different cells may reuse the same spectrum resource, and lead to severe inter-cell interference. To suppress the inter-cell interference and enhance the sum-rate of the cells reusing the same spectrum resource, power control algorithms are usually adopted \cite{HetNet_survey2}-\cite{HetNet_RA4}. Two conventional power control algorithms to maximize the sum-rate are \emph{weighted minimum mean square error} (WMMSE) algorithm \cite{WMMSE} and \emph{fractional programming} (FP) \cite{FP} algorithm. By assuming that the instantaneous \emph{global} (including both intra-cell and inter-cell) \emph{channel state information} (CSI) is available, both WMMSE and FP algorithms can be used to calculate power allocation policies for the \emph{access points} (APs) in different cells simultaneously.


It is known that, the solutions of conventional power control algorithms are generally sub-optimal. In fact, the solutions can be completely invalid when the coherence time of wireless channels is longer than the processing time, which is the total time of estimating channels, running power control algorithms, and updating transmit power. In a HetNet, the radio environment is highly dynamic and the coherence time of wireless channels is typically shorter than the processing time. As a result, the output solution of the conventional power control algorithms is typically invalid.

Motivated by the appealing performance of the machine learning in computer science field, e.g., computer vision and natural language processing, machine learning is recently advocated for wireless network designs. One typical application of the machine learning lies in the dynamic power control for the sum-rate maximization in wireless networks \cite{HetNet_DL} \cite{HetNet_DRL}. \cite{HetNet_DL} designs an \emph{deep learning} (DL) based algorithm to accelerate the power allocations in a general interference-channel scenario. By collecting a large number of global CSI sets, \cite{HetNet_DL} uses the WMMSE algorithm to generate power allocation labels. Then, \cite{HetNet_DL} trains a \emph{deep neural network} (DNN) with these global CSI sets and the corresponding power allocation labels. With the trained DNN, power allocation policies can be directly calculated by feeding the instantaneous global CSI. To avoid requiring the instantaneous global CSI meanwhile eliminate the computational cost of generating power allocation labels with the WMMSE algorithm, \cite{HetNet_DRL} considered a homogeneous network and assumed that neighboring transceivers (i.e., APs and users) can exchange their local information through certain cooperations. Then, \cite{HetNet_DRL} developed a \emph{deep reinforcement learning} (DRL) based algorithm, which optimizes power allocation policies in a trial-and-error manner and can converge to the performance of the WMMSE algorithm after sufficient trials.

\subsection{Contributions}

In this paper, we study the power control problem for the sum-rate maximization in a typical HetNet, in which multi-tier cells coexist by reusing the same spectrum band. Different from the single-tier scenario, both the maximum transmit power and coverage of each AP in the multi-tier scenario are typically heterogeneous. Main contributions of the paper are summarized as follows:
\begin{enumerate}


\item We exploit DRL to design a multi-agent power control algorithm in the HetNet. First, we establish a local DNN at each AP and treat each AP as an agent. The input and the output of each local DNN are the local state information and the adopted local action, i.e., transmit power of the corresponding AP, respectively. Then, we propose a novel \emph{multiple-actor-shared-critic} (MASC) method to train separately each local DNN in an online trial-and-error manner.

\item The MASC training method is composed of multiple actor DNNs and a shared critic DNN. In particular, we first establish an actor DNN in the core network for each local DNN, and the structure of each actor DNN is the same as the corresponding local DNN. Then, we establish a shared critic DNN in the core network for these actor DNNs. By feeding historical global information into the critic DNN, the output of the critic DNN can accurately evaluate whether the output (i.e., transmit power) of each actor DNN is good or not from an global view. By training each actor DNN with the evaluation of the critic DNN, the weight vector of each actor DNN can be updated towards the direction of the global optimum. The weight vector of each local DNN can be periodically replaced by that of the associated actor DNN until convergence.


\item The proposed algorithm has two main advantages compared with the existing power control algorithms. First, compared with \cite{WMMSE}, \cite{FP}, \cite{HetNet_DL}, and \cite{HetNet_DRL}, each AP in the proposed algorithm can independently control the transmit power and enhance the sum-rate based on only local state information, in the absence of instantaneous global CSI and any cooperation with other cells. Second, compared with \cite{HetNet_DRL}, \cite{DRL_0}, and \cite{DRL_1}, the reward function of each agent in the proposed algorithm is the transmission rate between the corresponding AP and its served user, avoiding particular reward function designs for each AP. This may ease the transfer of the proposed algorithm framework into more resource management problems in wireless communications.


\item By considering both two-layer and three-layer HetNets in simulations, we demonstrate that the proposed algorithm can rapidly converge to an average sum-rate higher than those of WMMSE and FP algorithms. Simulation results also reveal that the proposed algorithm outperforms WMMSE and FP algorithms in terms of the computational complexity.
\end{enumerate}

\subsection{Related literature}

DRL originates from RL, which has been widely used for the designs of wireless communications \cite{RL_0}-\cite{RL_5}, e.g., user/AP handoff, radio access technology selection, energy-efficient transmissions, user scheduling and resource allocation, spectrum sharing, and etc. In particular, RL estimates the long-term reward of each state-action pair and stores them into a two-dimensional table. For a given state, RL can choose the action subject to the maximum long-term reward to enhance the performance. It has been demonstrated that RL performs well in a decision-making scenario in which the size of the state and action spaces in the wireless system is relatively small. However, the effectiveness of RL diminishes when the state and action spaces become large.

To choose proper actions in the scenarios with large state and action spaces, DRL is proposed by properly integrating DL and RL \cite{DRL_nature}. In particular, by adopting the DNN which has a strong representation capability, DL has been applied in different areas of wireless communications \cite{DL_survey}, for example, power control \cite{DL_PC}, channel access \cite{DL_CA}, link scheduling \cite{DL_LS}. The basic idea of DRL is as follows: Instead of storing the long-term reward of each state-action pair in a tabular manner, DRL uses a DNN to represent the long-term reward as a function of the state and action. Thanks to the strong representation capability of the DNN, the long-term reward of any state-action pair can be properly approximated. The application of DRL in HetNets includes interference control among small cells \cite{DRL_0}, power control in a HetNet \cite{DRL_1}, resource allocation in V2V communications \cite{DRL_2}, caching policy optimization in content distribution networks \cite{DRL_3}, multiple access optimization in a HetNet \cite{DRL_4}, modulation and coding scheme selection in a cognitive HetNet \cite{DRL_5}, joint beamforming/power control/interference coordination in 5G networks \cite{DRL_6}, UAV navigation \cite{DRL_7}, and spectrum sharing \cite{DRL_8}. More applications of DRL in wireless communications can be found in \cite{DRL_survey}.


\subsection{Organizations of the paper}

The remainder of this paper is organized as follows. We provide the system model in Sec. II. In Sec. III, we provide the problem formulation and analysis. In Sec. IV, we give preliminaries by overviewing related DRL algorithms. In Sec. V, we elaborate the proposed power control algorithm. Simulation results are shown in Sec. VI. Finally, we conclude the paper in Sec. VII.

\section{System Model}
          \begin{figure}
            \centering
            \includegraphics[scale=0.8]{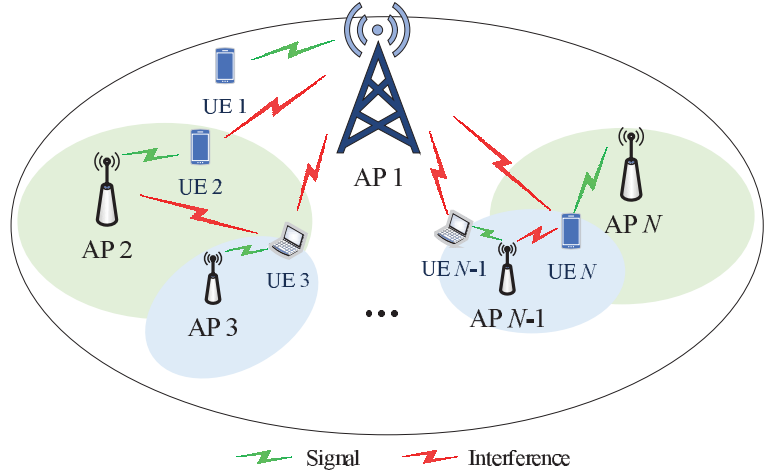}
            \caption{A general HetNet, in which multiple APs share the same spectrum band to serve the users within their coverages and may cause interference to each other.}
             \label{system_model}
        \end{figure}
As shown in Fig. \ref{system_model}, we consider a typical HetNet, in which multiple APs share the same spectrum band to serve users and may cause interference to each other. In particular, we denote an AP as AP $n$, $n\in \mathbb{N}=\{1, 2, \cdots, N\}$, where $N$ is the number of APs. Accordingly, we denote the user served by AP $n$ as \emph{user equipment} (UE) $n$. Next, we provide the channel model and signal transmission model, respectively.

\subsection{Channel model}
The channel between an AP and a UE has two components, i.e., large-scale attenuation (including path-loss and shadowing) and small-scale block Rayleigh fading. If we denote $\phi_{n,k}$ as the large-scale attenuation and denote $h_{n,k}$ as the small-scale block Rayleigh fading from AP $n$ to UE $k$, the corresponding channel gain is $g_{n,k}=\phi_{n,k}|h_{n,k}|^2$. In particular, the large-scale attenuation is highly related to the locations of the AP and the UE, and typically remains constant for a long time. The small-scale block Rayleigh fading remains constant in a single time slot and changes among different time slots.

According to \cite{Jake_model}, we adopt the Jake's model to represent the relationship between the small-scale Rayleigh fadings in two successive time slots, i.e.,
\begin{align}
h(t)=\rho h(t-1)+\omega,
\end{align}
where $\rho$ ($0\leq \rho \leq 1$) denotes the correlation coefficient of two successive small-scale Rayleigh fading realizations, $\omega$ is a random variable represented by a distribution $\omega \sim \mathcal{CN}(0, 1-\rho^2)$, and $h(0)$ is a random variable produced by a distribution $h(0) \sim \mathcal{CN}(0, 1)$. It should be noted that, the Jake's model can be reduced to the \emph{independent and identically distributed} (IID) channel model if $\rho$ is zero.

\subsection{Signal transmission model}

If we denote $x_n(t)$ as the downlink signal from AP $n$ to UE $n$ with unit power in time slot $t$, the received signal at UE $n$ is
\begin{align}
    y_n(t)= \sqrt{p_n(t)\phi_{n,n}}h_{n,n}(t)x_n(t)+\sum_{k \in \mathbb{N},k\neq n}\sqrt{p_k(t)\phi_{k,n}}h_{k,n}(t)x_k(t)+\delta_n(t),
    \label{y_n}
\end{align}

where $p_n(t)$ is the transmit power of AP $n$, $\delta_n(t)$ is the noise at UE $n$ with power $\sigma^2$, the first term on the right side is the received requiring signal from AP $n$, and the second term on the right side is the received interference from other APs. By considering that all the downlink transmissions from APs to UEs are synchronized, the \emph{signal to interference and noise ratio} (SINR) at UE $n$ can be written as
\begin{align}
    \gamma_n(t)= \frac{p_n(t)g_{n,n}(t)}{\sum_{k \in \mathbb{N},k\neq n}p_k(t)g_{k,n}(t)+\sigma^2},
    \label{SNR_n}
\end{align}
Accordingly, the downlink transmission rate (in bps) from AP $n$ to UE $n$ is
\begin{align}
    r_n(t)= B\log\left(1+\gamma_n(t)\right),
    \label{rate_n}
\end{align}
where $B$ is the bandwidth of the downlink transmission.

\section{Problem description and analysis}

Our goal is to optimize the transmit power of all APs to maximize the sum-rate of all the downlink transmissions. Then, we can formulate the optimization problem as
 \begin{align}
\nonumber \ \ \underset{p_n, \forall n \in \mathbb{N}}{\max} \ \ & R(t)=\sum_{n=1}^{N} r_n(t)\\
{{\rm{s}}{\rm{.t}}{\rm{.}}} \ \ & 0\leq p_n(t) \leq p_{n,\text{max}}, \forall n \in \mathbb{N},
\end{align}
where $p_{n,\text{max}}$ is the maximum transmit power constraint of AP $n$. Note that the APs of different cells (e.g., macro-cell base station, pico-cell base station, and femto-cell base station) typically have distinct maximum transmit power constraints.

Since wireless channels are typically dynamic, the optimal transmit power maximizing the sum-rates among distinct time slots differs a lot. In other words, the optimal transmit power should be determined at the beginning of each time slot to guarantee the optimality. Nevertheless, there exist two main challenges to determine the optimal transmit power of APs. First, according to \cite{NP_Hard}, this problem is generally NP-hard and it is difficult to find the optimal solution. Second, the optimal transmit power should be determined at the beginning of each time slot and it is demanding to find the optimal solution (if possible) in such a short time period.

As mentioned above, conventional power control algorithms (e.g., WMMSE algorithm and FP algorithm) can output sub-optimal solutions for this problem by implicitly assuming a quasi-static radio environment, in which wireless channels change slowly. For the dynamic radio environment, DL \cite{HetNet_DL} and DRL \cite{HetNet_DRL} are recently adopted to solve the problem with sub-optimal solutions, by assuming the availability of the instantaneous global CSI or the cooperations among neighboring APs. In fact, these algorithms are inapplicable in the considered scenario of this paper due to the following two main constraints:
\begin{itemize}
\item Instantaneous global CSI is unavailable.

\item Neighboring APs are not willing to or even cannot cooperate with each other.
\end{itemize}

In the rest of the paper, we will first provide preliminaries by overviewing related DRL algorithms and then develop a DRL based multi-agent power control algorithm to solve the above problem.

\section{Preliminaries}
In this section, we provide an overview of two DRL algorithms, i.e., \emph{Deep Q-network} (DQN) and \emph{Deep deterministic policy gradient} (DDPG), both of which are important to develop the power control algorithm in this paper. In general, DRL algorithms mimic the human being to solve a successive decision-making problem via trial-and-errors. By observing the current environment state $s$ ($s\in \textbf{\emph{S}}$) and adopting an action $a$ ($a \in \textbf{\emph{A}}$) based on a policy $\pi$, the DRL agent can obtain an immediate reward $r$ and observes a new environment state $s'$. By repeating this process and treating each tuple $\{s, a, r, s'\}$ as an experience, the DRL agent can continuously learn the optimal policy from the experiences to maximize the long-term reward.


\subsection{Deep Q-network (DQN) \cite{DRL_nature}}

DQN establishes a DNN $Q(s,a;\theta)$ with weight vector $\theta$ to represent the expected cumulative discounted (long-term) reward by executing the action $a$ in the environmental state $s$. Then, $Q(s,a;\theta)$ can be rewritten in a recursive form (Bellman equation) as
\begin{align}
Q(s,a;\theta) = r(s, a) + \eta \sum\limits_{s' \in \textbf{\emph{S}}} \sum\limits_{a' \in \textbf{\emph{A}}} {{p_{s,s'}}(a)} Q(s', a';\theta),
\label{DRL_1}
\end{align}
where $r(s, a)$ is the immediate reward by executing the action $a$ in the environmental state $s$, $\eta \in [0, 1]$ is the discount factor representing the discounted impact of future rewards, and ${p_{s,s'}}(a)$ is the transition probability of the environment state from $s$ to $s'$ by executing the action $a$. The DRL agent aims to find the optimal weight vector $\theta^*$ to maximize the long-term reward for each state-action pair. With the optimal weight vector $\theta^*$, (\ref{DRL_1}) can be rewritten as
\begin{align}
Q(s,a;\theta^*) = r(s,a) + \eta \sum\limits_{s' \in \textbf{\emph{S}}} {{p_{s,s'}}(a)\mathop {\max }\limits_{a' \in \textbf{\emph{A}}} } \ Q(s', a';\theta^*).
\label{DRL_2}
\end{align}
Accordingly, the optimal action policy is
\begin{align}
{\pi ^*}(s) = \mathop {\arg \max }\limits_{a \in \textbf{\emph{A}}} \left[ {Q(s,a;\theta^*)} \right], \ \forall \ s\in \textbf{\emph{S}}.
\end{align}

However, it is challenging to directly obtain the optimal $\theta^*$ from (\ref{DRL_2}) since the transition probability ${p_{s,s'}}(a)$ is typically unknown to the DRL agent. Then, DRL agent updates $\theta$ in an iterative manner. On the one hand, to balance exploitation and exploration, the DRL agent adopts an $\epsilon$-greedy algorithm to choose an action for each environmental state: for a given state $s$, the DRL agent executes the action $a=\arg \max_{a\in  \textbf{\emph{A}}}Q(s,a;\theta)$ with the probability $1-\epsilon$, and randomly executes an action with the probability $\epsilon$. On the other hand, by executing actions with $\epsilon$-greedy algorithm, DRL agent can continuously accumulate experience $e=\{s,a,r,s'\}$ and store it in an experience replay buffer in a \emph{first-in-first-out} (FIFO) fashion. After sampling a mini-batch of experiences $\mathcal{E}$ with length $D$ ($\mathcal{E}=\{e_1, e_2,\cdots, e_D\}$) from the experience replay buffer, the DRL agent can update $\theta$ by adopting a stochastic gradient decent (SGD) method to minimize the expected prediction error (loss function) of the sampled experiences, i.e.,
     \begin{align}
\mathbb{L}(\theta ) = \frac{1}{D}\sum_{\mathcal{E}}{{{\left[r + \eta \mathop {\max }\limits_{a' \in \textbf{\emph{A}}} \ Q^-(s^{'},a';\theta^-) - Q(s,a;\theta)\right]}^2}} ,
\label{loss_function}
\end{align}
where $Q^-(s,a;\theta^-)$ is the target DNN and is established with the same structure as $Q(s,a;\theta)$. In particular, the weight vector $\theta^-$ is updated by $\theta$ periodically to stabilize the training of $Q(s,a;\theta)$.



\subsection{Deep deterministic policy gradient (DDPG) \cite{DDPG}}
It should be noted that DQN needs to estimate the long-term reward for each state-action pair to obtain the optimal policy. When the action space is continuous, DQN is inapplicable and DDPG can be an alternative. DDPG has an actor-critic architecture, which includes an actor DNN $\mu(s;\theta_{\mu})$ and a critic DNN $Q(s,a;\theta_{Q})$. In particular, the actor DNN $\mu(s;\theta_{\mu})$ is the policy network and is responsible to output a deterministic action $a$ from a continuous action space in an environment state $s$, and the critic DNN is the value-function network (similar to DQN) and is responsible to estimate the long-term reward when executing the action $a$ in the environment state $s$. Similar to DQN, to stabilize the training of actor DNN and critic DNN, DDPG establishes an target actor DNN $\mu^-(s;\theta_{\mu}^-)$ and a target critic DNN $Q^-(s,a;\theta_{Q}^-)$, respectively.

Similar to the DQN, the DDPG agent samples experiences from the experience replay buffer to train the actor DNN and critic DNN. Nevertheless, the DDPG accumulates experiences in a unique way. In particular, the executed action in each time slot is a noisy version of the output of the actor DNN, i.e., $a=\mu(s;\theta_{\mu})+\zeta$, where $\zeta$ is a random variable and guarantees continuously exploring the action space around the output of the actor DNN. Note that, the epsilon-greedy method is usually used for action selections in discrete action space scenarios. When the action space is continuous, the noisy version of the output of the actor DNN is preferable for action selections \cite{DDPG} \cite{CCDRL}.

The training procedure of the critic DNN is similar to that of DQN: by sampling a mini-batch of experiences $\mathcal{E}$ with length $D$ from the experience replay buffer, the DDPG agent can update $\theta_{Q}$ by adopting a SGD method to minimize the expected prediction error of the sampled experiences, i.e.,
\begin{align}
\mathbb{L}(\theta_{Q}) = \frac{1}{D}\sum_{\mathcal{E}}{{{\left[r + \eta \mathop {\max }\limits_{a' \in \textbf{\emph{A}}} \ Q^-(s_i^{'},a';\theta_{Q}^-)- Q(s,a;\theta_{Q})\right]}^2}}.
 \end{align}


Then, the DDPG agent adopts a soft update method to update the weight vector $\theta_{Q}^-$ of the target critic DNN, i.e., $\theta_{Q}^- \gets \tau \theta_{Q} +(1-\tau)\theta_{Q}^-$,
where $\tau \in [0, 1]$ is the learning rate of the target DNN.


The goal of training the actor DNN is to maximize the expected value-function $Q\left(s,\mu(s;\theta_{\mu});\theta_{Q}\right)$ in terms of the environment state $s$, i.e., $J(\theta_{\mu})=\mathbb{E}_{s}\left[Q\left(s,\mu(s;\theta_{\mu});\theta_{Q}\right)\right]$. Then, the weight vector $\theta_{\mu}$ of the actor DNN can be updated in the sampled gradient direction of $J(\theta_{\mu})$, i.e.,
\begin{align}
\nabla_{\theta_{\mu}} J(\theta_{\mu})
\approx \frac{1}{D}\sum_{\mathcal{E} }\nabla_{a}Q\left(s,a;\theta_{Q}\right)|_{a=\mu(s;\theta_{\mu})} \nabla_{\theta_{\mu}}\mu(s;\theta_{\mu}).
\label{actor_update_direction}
\end{align}



Similar to the target critic DNN, the DDPG agent updates the weight vector of the target actor DNN by $\theta_{\mu}^- \gets \tau \theta_{\mu} +(1-\tau)\theta_{\mu}^-$.

\section{DRL for multi-agent power control algorithm}

In this section, we exploit DRL to design a multi-agent power control algorithm for the APs in the HetNet. In the following, we will first introduce the algorithm framework and then elaborate the algorithm design.

\subsection{Algorithm framework}

It is known that the optimal transmit power of APs is highly related to the instantaneous global CSI. But the instantaneous global CSI is unavailable to APs. Thus, it is impossible for each AP to optimize the transmit power through conventional power control algorithms, e.g., WMMSE algorithm or FP algorithm. From \cite{HetNet_DL} and \cite{HetNet_DRL}, the historical wireless data (e.g., global CSI, transmit power of APs, the mutual interference, and achieved sum-rates) of the whole network contains useful information that can be utilized to optimize the transmit power of APs. Thus, we aim to leverage DRL to develop an intelligent power control algorithm that can fully utilize the historical wireless data of the whole network. From the aspect of practical implementations, the intelligent power control algorithm should have two basic functionalities:
\begin{itemize}
\item \emph{Functionality I}: Each AP can use the algorithm to complete the optimization of the transmit power at the beginning of the each time slot, in order to guarantee the timeliness of the optimization.

%
\item \emph{Functionality II}: Each AP can independently optimize the transmit power to enhance the sum-rate with only local observations, in the absence of the global CSI and any cooperations among APs.
\end{itemize}

          \begin{figure}
            \centering
            \includegraphics[scale=0.7]{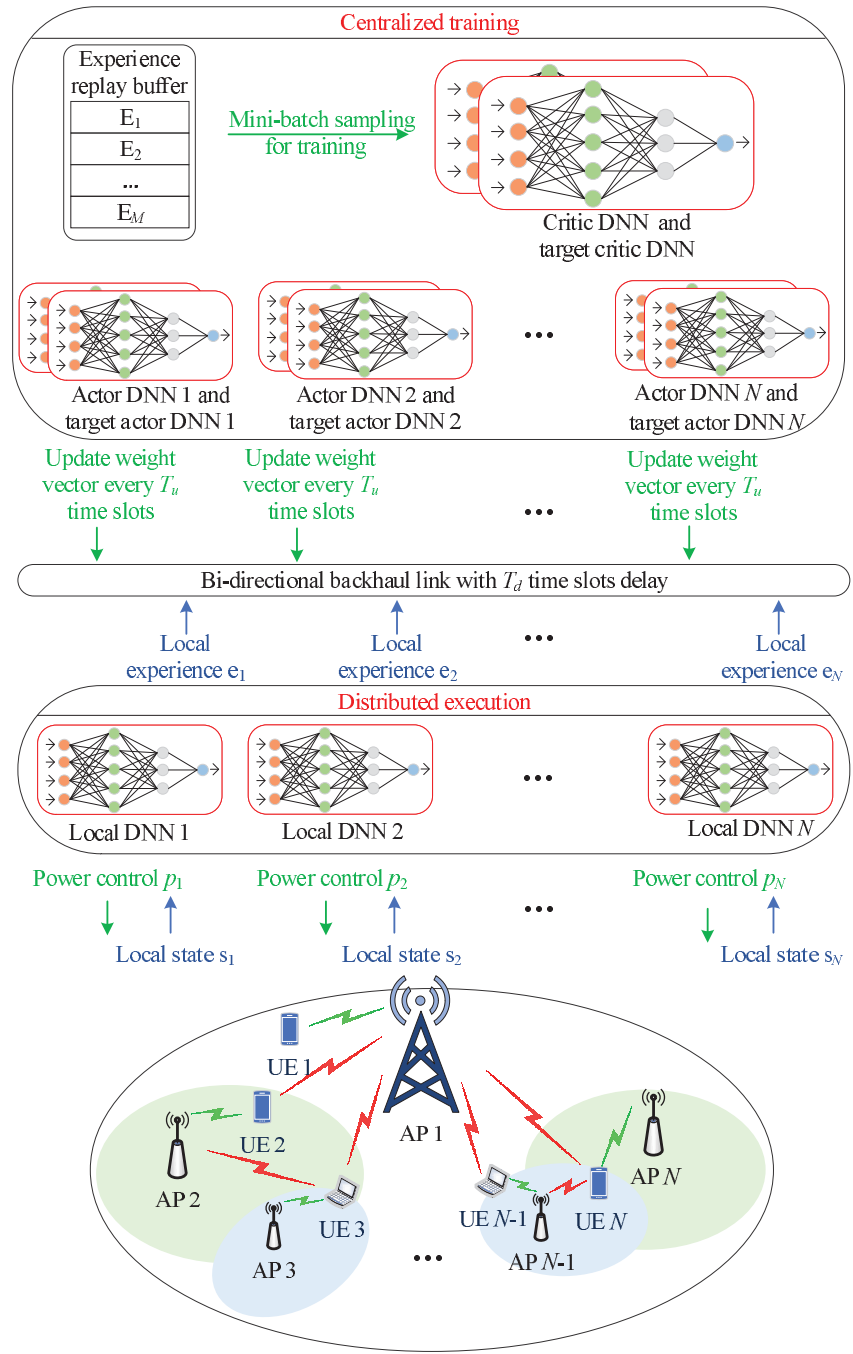}
            \caption{Proposed algorithm framework.}
             \label{Algorithm_framework}
        \end{figure}

To realize Functionality I, we adopt a centralized-training-distributed-execution architecture as the basic algorithm framework as shown in Fig. \ref{Algorithm_framework}. To be specific, a local DNN is established at each AP, and the input and the output of each local DNN are the local state information and the adopted local action, i.e., transmit power of the corresponding AP, respectively. In this way, each AP can feed local observations into the local DNN to calculate the transmit power in a real-time fashion. The weight vector of each local DNN is trained in the core network, which has redundant historical wireless data of the whole network. Since there are $N$ APs, we denote the corresponding $N$ local DNNs as $\mu_n^{\text{(L)}}\left(s_n;\theta_n^{\text{(L)}}\right)$ ($n\in \mathbb{N}$), in which $\theta_n^{\text{(L)}}$ is the weight vector of the local DNN $n$ and $s_n$ is the local state of AP $n$.

To realize Functionality II, we develop a MASC training method based on the DDPG to update the weight vectors of local DNNs. To be specific, as shown in Fig. \ref{Algorithm_framework}, $N$ actor DNNs together with $N$ target actor DNNs are established in the core network to associate with $N$ local DNNs, respectively. Each actor DNN has the same structure as the associated local DNN, such that the trained weight vector of the actor DNN can be used to update the associated local DNN. Meanwhile, a shared critic DNN together with the corresponding target critic DNN is established in the core network to guide the training of $N$ actor DNNs. The inputs of the critic DNN include the global state information and the adopted global action, i.e., the transmit power of each AP, and the output of the critic DNN is the long-term sum-rate of the global state-action pair. It should be noted that, there are two main benefits to include the global state and global action in the input of the critic DNN. First, by doing this, the non-stationary radio environment issue the critic DNN faces is only caused by the time-variant nature of wireless channels. This is the major difference compared with the case, in which single-agent algorithm is directly applied to solve a multi-agent problem and the non-stationary issue is caused by both the time-variant nature of wireless channels and unknown actions of other agents. The adverse impact of the non-stationary radio environment issue on the system performance can thus be reduced. Second, by doing this, we can train the critic DNN with historical global state-action pairs together with the achieved sum-rates in the core network, such that the critic DNN has a global view of the relationship between the global state-action pairs and the long-term sum-rate. Then, the critic DNN can evaluate whether the output of an actor DNN is good or not in terms of the long-term sum-rate. By training each actor DNN with the evaluation of the critic DNN, the weight vector of each actor DNN can be updated towards the direction of the global optimum. To this end, the weight vector of each local DNN can be periodically replaced by that of the associated actor DNN until convergence.



We denote $N$ actor DNNs as $\mu_n^{(\text{a})}\left(s_n;\theta_n^{(\text{a})}\right)$, ($n\in \mathbb{N}$), where $\theta_n^{(\text{a})}$ is the weight vector of actor DNN $n$. Accordingly, we denote $N$ target actor DNNs as $\mu_n^{(\text{a-})}\left(s_n;\theta_n^{(\text{a-})}\right)$, ($n\in \mathbb{N}$), where $\theta_n^{(\text{a-})}$ is the corresponding weight vector. Then, we denote respectively the critic DNN and the target critic DNN as $Q\left(s_1, \cdots, s_N, s_{\text{o}}, a_1, \cdots, a_N;\theta^{(\text{c})}\right)$ and $Q^-\left(s_1, \cdots, s_N, s_{\text{o}}, a_1, \cdots, a_N;\theta^{(\text{c-})}\right)$, where $\{s_1, \cdots, s_N, s_{\text{o}}\}$ is referred to as the global state of the whole network including all the local states $s_n$ ($n\in \mathbb{N}$) and other global state $s_{\text{o}}$ of the whole network, $\{a_1, \cdots, a_N\}$ is referred to as global action of the whole network including all the actions $a_n$ ($ \forall \ n\in \mathbb{N}$), $\theta^{(\text{c})}$ and $\theta^{(\text{c-})}$ are the corresponding weight vectors.

Next, we detail the experience accumulation procedure followed by the MASC training method.

\subsubsection{Experience accumulation}
At the beginning of time slot $t$, AP $n$ ($\forall \ n\in \mathbb{N}$) observes a local state $s_n$ and optimizes the action (i.e., transmit power) as $a_n=\mu_n^{\text{(L)}}\left(s_n;\theta_n^{\text{(L)}}\right)+\zeta$, where the action noise $\zeta$ is a random variable and guarantees continuously exploring the action space around the output of the local DNN. By executing the action (i.e., transmit power) within this time slot, each AP can obtain a reward (i.e., transmission rate) $r_n$ in the end of time slot $t$. At the beginning of the next time slot, AP $n$, ($\forall \ n\in \mathbb{N}$), observes a new local state $s'_n$. To this end, AP $n$, ($\forall \ n\in \mathbb{N}$), can obtain a local experience $e_n=\{s_n, a_n, r_n, s_n^{'}\}$ and meanwhile upload it to the core network via a bi-directional backhaul link with $T_d$ time slots delay. Upon receiving $e_n$, ($ \forall \ n\in \mathbb{N}$), a global experience is constructed as $E=\{s_1,\cdots, s_N, s_{\text{o}}, a_1, \cdots, a_N, R, s_1^{'}, \cdots, s_N^{'}, s_{\text{o}^{'}}\}$, where $R=\sum_{n=1}^{N}r_n$ is the global reward of the whole network, and each global experience will be stored in the experience replay buffer, which has the capacity of $M$ and works in an FIFO fashion. By repeating this procedure, the experience replay buffer can continuously accumulate new global experiences.

\subsubsection{MASC training method}
To train the critic DNN, a mini-batch of experiences $\mathcal{E}$ can be sampled from the experience replay buffer in each time slot. Then, the SGD method can be adopted to minimize the expected prediction error (loss function) of the sampled experiences, i.e.,
     \begin{align}
\!\!\mathbb{L}(\theta^{(\text{c})})
\!\!=\!\! \frac{1}{D}\!\sum_{\mathcal{E}}{{{[y^\text{Tar} \!\!-\!\! Q(s_1,\cdots, s_N,s_{\text{o}}, a_1, \cdots, a_N;\theta^{(\text{c})})]}^2}} ,
\label{loss_function_critic}
\end{align}
where $y^{\text{Tar}}$ can be calculated by
\begin{align}
{y^\text{Tar}}
=\sum_{n=1}^{N}r_n + \eta \mathop {\max }\limits_{a_n^{'} \in \textbf{\emph{A}}} Q^-(s_1^{'}, \cdots, s_N^{'}, s_{\text{o}}, a_1^{'}, \cdots, a_N^{'};\theta^{(\text{c-})}).
\label{trained_output_critic}
\end{align}
Then, a soft update method is adopted to update the weight vector $\theta^{(\text{c-})}$ of the critic target DNN, i.e.,
\begin{align}
\theta^{(\text{c-})} \gets  \tau^{(\text{c})} \theta^{(\text{c})} +(1-\tau^{(\text{c})})\theta^{(\text{c-})},
\label{soft_update_critic}
\end{align}
where $\tau^{(\text{c})} \in [0, 1]$ is the learning rate of the target critic DNN.

Since each AP aims to optimize the sum-rate of whole network, the training of the actor DNN $n$, ($ \forall \ n\in \mathbb{N}$), can be designed to maximize the expected long-term global reward $J(\theta_1^{\text{(a)}}, \cdots, \theta_N^{\text{(a)}})$, which is defined as the expectation of $Q\left(s_1, \cdots, s_N, s_{\text{o}}, \mu_1^{\text(a)}(s_1;\theta_{1}^{\text{(a)}}), \cdots, \mu_N^{\text(a)}(s_N;\theta_{N}^{\text{(a)}});\theta^{\text{(c)}}\right)$ in terms of the global state $\{s_1, \cdots, s_N, s_{\text{o}}\}$, i.e.,
\begin{align}
J(\theta_1^{\text{(a)}}, \cdots, \theta_N^{\text{(a)}})
=\mathbb{E}_{s_1, \cdots, s_N, s_{\text{o}}}\left[Q\left(s_1, \cdots, s_N, s_{\text{o}}, \mu_1^{\text(a)}(s_1;\theta_{1}^{\text{(a)}}), \cdots, \mu_N^{\text(a)}(s_N;\theta_{N}^{\text{(a)}});\theta^{\text{(c)}}\right)\right].
\label{Global_objective}
\end{align}


By taking the partial derivation of $J(\theta_1^{\text{(a)}}, \cdots, \theta_N^{\text{(a)}})$ with respect to $\theta_n^{\text{(a)}}$, we have
\begin{align}
\nabla_{\theta_{n}^{\text{(a)}}} J(\theta_1^{\text{(a)}}, \cdots, \theta_N^{\text{(a)}})
\approx\frac{1}{D}\sum_{\mathcal{E} }\nabla_{a_n}Q\left(s_1, \cdots, s_N, s_{\text{o}}, a_1, \cdots, a_N;\theta^{\text{(c)}}\right)|_{a_n=\mu(s_n;\theta_{n}^{\text{(a)}})} \nabla_{\theta_{n}^{\text{(a)}}}\mu_n^{\text{(a)}}(s_n;\theta_{n}^{\text{(a)}}).
\label{actor_n_update_direction}
\end{align}
Then, $\theta_{n}^{\text{(a)}}$ can be updated in the direction of $\nabla_{\theta_{n}^{\text{(a)}}} J(\theta_1^{\text{(a)}}, \cdots, \theta_N^{\text{(a)}})$, which is the direction with the maximum likelihood to increase $J(\theta_1^{\text{(a)}}, \cdots, \theta_N^{\text{(a)}})$.

Similar to the target critic DNN, a soft update method is adopted to adjust the weight vector $\theta_n^{(\text{a-})}$ of the target actor DNN, i.e.,
\begin{align}
\theta_n^{(\text{a-})} \gets \tau_n^{(\text{a})} \theta_n^{(\text{a})} +(1-\tau_n^{(\text{a})})\theta_n^{(\text{a-})},
\label{soft_update_actor}
\end{align}
where $\tau_n^{\text{(a)}} \in [0, 1]$ is the learning rate of the corresponding target actor DNN.

\subsection{Algorithm designs}
In this part, we first design the experience and the structure for each actor DNN and local DNN. Then, we design the global experience and the structure of the critic DNN. Finally, we elaborate the algorithm followed by some related discussions.

          \begin{figure}
            \centering
            \includegraphics[scale=0.7]{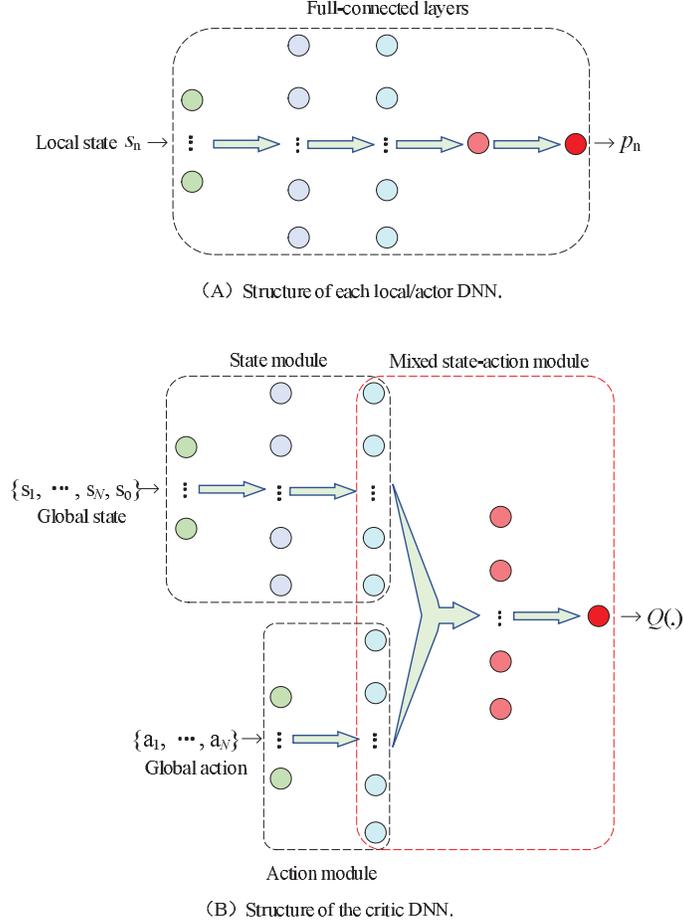}
            \caption{The structures of each actor DNN and the critic DNN, in which the arrows indicate the direction of the data flow. Each dotted box contains a certain number of hidden layers.}
             \label{DNN_structures}
        \end{figure}

\subsubsection{Experience of actor DNNs}
The local information at AP $n$ can be divided into historical local information in previous time slots and instantaneous local information at the beginning of the current time slot. Historical local information includes the channel gain between AP $n$ and UE $n$, the transmit power of AP $n$, the sum-interference from AP $k$ ($k \in \mathbb{N}, k\neq n$), the received SINR, and the corresponding transmission rate. Instantaneous local information includes the channel gain between AP $n$ and UE $n$, and the sum-interference from AP $k$ ($k \in \mathbb{N}, k\neq n$). It should be noted that, the sum-interference from AP $k$ ($k \in \mathbb{N}, k\neq n$) at the beginning of the current time slot is generated as follows: At the beginning of the current time slot, the new transmit power of each AP has not been determined, and each AP still uses the transmit power of the previous time slot, although the CSI of the whole network has changed. Thus, state $s_n$ in time slot $t$ is designed as
\begin{align}\nonumber
\!s_n(t)=&\left\{g_{n,n}(t\!-\!1), p_n(t\!-\!1), \!\!\!\!\sum_{k \in \mathbb{N} ,k\neq n} \!\!\!\!p_k(t\!-\!1)g_{k,n}(t\!-\!1),\right.\\
 &\left.\gamma_n(t\!-\!1), r_n(t\!-\!1), g_{n, n}(t), \!\!\!\!\sum_{k \in \mathbb{N} ,k\neq n}\!\!\!\! p_k(t\!-\!1)g_{k,n}(t) \right\}.
\label{actor_n_state}
\end{align}


Besides, action $a_n(t)$ and reward $r_n(t)$ can be respectively designed as the transmit power $a_n(t)=p_n(t)$ and the corresponding achievable rate calculated by (\ref{rate_n}). Consequently, the experience $e_n(t)$ can be constructed as
\begin{align}
e_n(t)=\{s_n(t-1), a_n(t-1), r_n(t-1), s_n(t)\}.
\label{actor_n_experience}
\end{align}

\subsubsection{Structure of local/actor DNNs}
The designed structure of the local/actor DNN includes five full-connected layers as illustrated in Fig. \ref{DNN_structures}-(A). In particular, the first layer is the input layer for $s_n$ and has $L_1^{\text{(a)}}=7$ neurons corresponding to seven elements in $s_n$. The second layer and the third layer have respectively $L_2^{\text{(a)}}$ and $L_3^{\text{(a)}}$ neurons. The forth layer has $L_4^{\text{(a)}}=1$ neuron with the sigmoid activation function, which outputs a value between zero and one. The forth layer has $L_5^{\text{(a)}}=1$ neuron, which scales linearly the value from the forth layer to a value between zero and $p_{n,max}$. With this structure, each local/actor DNN can take the local state as the input and output a transmit power satisfying the maximum transmit power constraint. In summary, there are $L_2^{\text{(a)}}+L_3^{\text{(a)}}+9$ neurons in each local/actor DNN.

\subsubsection{Global experience}
By considering $T_d$ time slots delay for the core network to obtain the local information of APs, we design the global experience in time slot $t$ as
\begin{align}\nonumber
E(t)=&\left\{s_1(t-1-T_d),\cdots, s_N(t-1-T_d), s_{\text{o}}(t-1-T_d),\right. \\ \nonumber
& \left. a_1(t-1-T_d), \cdots, a_N(t-1-T_d), R(t-1-T_d),\right. \\
& \left. s_1(t-T_d), \cdots, s_N(t-T_d), s_{\text{o}}(t-T_d)\right\}.
\label{global_experience}
\end{align}
In particular, $s_n(t-1-T_d)$, $s_n(t-T_d)$, $a_n(t-1-T_d)$, and $a_n(t-1-T_d)$ ($ \forall \ n \in \mathbb{N}$) can be directly obtained from $e_n(t-T_d)$ ($ \forall \ n \in \mathbb{N}$), and $R(t-1-T_d)=\sum_{n \in \mathbb{N}} r_n(t-1-T_d)$ can be directly calculated with the local reward $r_n(t-1-T_d)$ in $e_n(t-T_d)$ ($ \forall \ n \in \mathbb{N}$). Here, we construct respectively $s_{\text{o}}(t-1-T_d)$  and $s_{\text{o}}(t-T_d)$ as $s_{\text{o}}(t-1-T_d)=G(t-1-T_d)$ and $s_{\text{o}}(t-T_d)=G(t-T_d)$, where $G(t-1-T_d)$ and $G(t-T_d)$ are the channel gain matrixes of the whole network in time slot $t-1-T_d$ and time slot $t-T_d$, respectively. Since the channel gains $g_{n,n}(t-1-T_d)$ and $g_{n,n}(t-T_d)$ ($\forall \ n \in \mathbb{N}$) are available in $s_n(t-1-T_d)$ and $s_n(t-T_d)$, we focus on the derivation of the interference channel gains $g_{n,k}(t-1-T_d)$ and $g_{n,k}(t-T_d)$ ($\forall \ n \in \mathbb{N}, \ k \in \mathbb{N}, \ n\neq k$). Here, we take the derivation of $g_{n,k}(t-T_d)$ ($\forall \ n \in \mathbb{N}, \ k \in \mathbb{N}, \ n\neq k$) as an example. At the beginning of time slot $t-T_d$, APs transmit orthogonal pilots to the corresponding UEs for channel and interference estimations, and UE $n$ can measure locally $p_k(t-1-T_d)g_{k,n}(t-T_d)$ from AP $k$ ($\forall \ k\in \mathbb{N}, \ k\neq n$). Then, UE $n$ can deliver the auxiliary information $o_n(t-T_d)=\{p_k(t-1-T_d)g_{k,n}(t-T_d), \forall \ k\in \mathbb{N}, \ k\neq n\}$ together with local state $s_n(t-T_d)$ to the local DNN at the beginning of time slot $t-T_d$. Then, by collecting simultaneously local experience $e_n(t-T_d)$ and the auxiliary information $o_n(t-T_d)$ from local DNN $n$ ($\forall \ n\in \mathbb{N}$), the core network can calculate each interference channel gain $g_{n,k}(t-T_d)$ ($\forall \ n \in \mathbb{N}, \ k \in \mathbb{N}, \ n\neq k$) with $p_n(t-1-T_d)g_{n,k}(t-T_d)$ in $o_k(t-T_d)$ and $p_n(t-1-T_d)$ in $s_n(t-T_d)$. In this way, the interference channel gains $g_{n,k}(t-T_d)$ ($\forall \ n \in \mathbb{N}, \ k \in \mathbb{N}, \ n\neq k$) can be obtained to construct $G(t-T_d)$. Following a similar procedure, the core network can construct the channel gain of the whole network in each time slot, including $G(t-1-T_d)$.

\subsubsection{Structure of the critic DNN}

The designed structure of critic DNN are illustrated in Fig. \ref{DNN_structures}-(B) including three modules, i.e., a state module, an action module, and a mixed state-action module. Each module contains several full-connected layers. For the state module, there are three full-connected layers. The first layer is the input layer for the global state $\{s_1, \cdots, s_N, s_{\text{o}}\}$. Since each $s_n$ ($n\in \ \mathbb{N}$) has seven elements and $s_{\text{o}}$ has $N^2$ elements, the first layer has $L_{1}^{\text{(S)}}=7N+N^2$ neurons. The second layer and the third layer of the state module have respectively $L_2^{\text{(S)}}$ and $L_3^{\text{(S)}}$ neurons. For the action module, there are two full-connected layers. The first layer of the action module is the input layer for the global action $\{a_1, \cdots, a_N\}$. Since each $a_n$ ($n\in \ \mathbb{N}$) is a one-dimension scalar, the first layer of the state module has $L_{1}^{\text{(A)}}=N$ neurons. The second layer of the action module has $L_{2}^{\text{(A)}}$ neurons. For the mixed state-action module, there are three full-connected layers. The first layer of the mixed state-action module is formulated by concatenating the last layers of the state module and the action module, and thus has $L_{1}^{\text{(M)}}=L_3^{\text{(S)}}+L_2^{\text{(A)}}$ neurons. The second layer of the mixed state-action module has $L_{2}^{\text{(M)}}$ neurons. The third layer of the mixed state-action module has one neuron, which outputs the long-term reward $Q^{\text{(c)}}(s_1,\cdots, s_N,s_{\text{o}}, a_1, \cdots, a_N;\theta^{(\text{c})})$. In summary, there are $N^2+8N+1+L_2^{\text{(S)}}+L_3^{\text{(S)}}+L_{2}^{\text{(A)}}+L_{2}^{\text{(M)}}$ neurons in the critic DNN.

\subsubsection{Proposed algorithm}

The proposed algorithm is illustrated in Algorithm 1, which includes three stages, i.e., Initializations, Random experience accumulation, and Repeat.

In the stage of Initializations, $N$ local DNNs, $N$ actor DNNs, $N$ target actor DNNs, a critic DNN, and a target critic DNN need to be properly constructed and initialized. In particular, local DNN $\mu_{n}^{\text{(L)}}(s_n;\theta_{n}^{\text{(L)}})$ ($\forall \ n \in \ \mathbb{N}$) is established at AP $n$ by adopting the structure in Fig. \ref{DNN_structures}-(A), actor DNN $\mu_{n}^{\text{(a)}}(s_n;\theta_{n}^{\text{(a)}})$ with and the corresponding target actor DNN $\mu_{n}^{\text{(a-)}}(s_n;\theta_{n}^{\text{(a-)}})$ are established in the core network to associate with local DNN $n$ by adopting the structure in Fig. \ref{DNN_structures}-(A), meanwhile critic DNN $Q^{\text{(c)}}(s_1,\cdots, s_N,s_{\text{o}}, a_1, \cdots, a_N;\theta^{(\text{c})})$ and the corresponding target critic DNN $Q^{\text{c-}}(s_1,\cdots, s_N,s_{\text{o}}, a_1, \cdots, a_N;\theta^{(\text{c-})})$ are established in the core network by adopting the structure in Fig. \ref{DNN_structures}-(B). Then, $\theta_{n}^{\text{(L)}}$, $\theta_{n}^{\text{(a)}}$, and $\theta^{(\text{c})}$ are randomly initialized, $\theta_{n}^{\text{(a-)}}$ and $\theta^{(\text{c-})}$ are initialized with $\theta_{n}^{\text{(a)}}$ and $\theta^{(\text{c})}$, respectively.

         \begin{figure}
            \centering
            \includegraphics[scale=0.55]{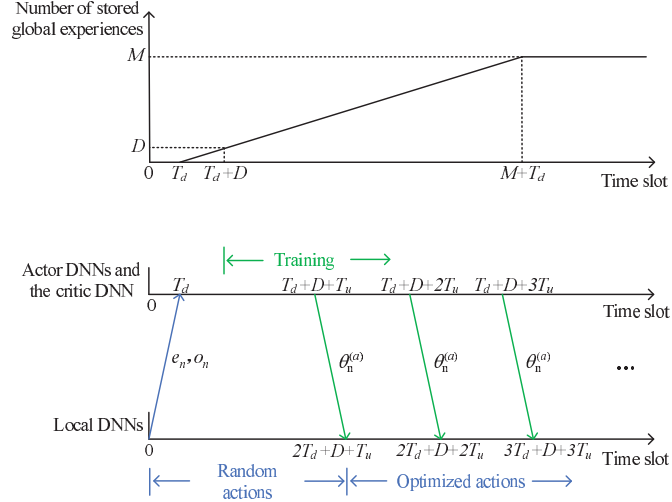}
            \caption{Diagram of the proposed algorithm.}
             \label{Algorithm_flowchart}
        \end{figure}

In the stage of Random experience accumulation ($t\leq 2T_d+D+T_u$), UE $n$ ($\forall \ n \in \ \mathbb{N}$) observes local state $s_n(t)$ and auxiliary information $o_n(t)$, and transmit them to AP $n$ at the beginning of time slot $t$. AP $n$ chooses a random action (i.e., transmit power) and meanwhile uploads local experience $e_n(t)$ and $o_n(t)$ to the core network through the bi-directional backhaul link with $T_d$ time slots delay. After collecting all the local experiences and the corresponding auxiliary information from $N$ APs, the core network constructs a global experience $E(t)$ and stores it into the memory relay buffer with the capacity $M$. As illustrated in Fig. \ref{Algorithm_flowchart}, at the beginning of time slot $T_d+D$, the core network has $D$ global experiences in the memory replay buffer. From time slot $T_d+D$, the core network begins to sample a mini-batch of experiences $\mathcal{E}$ with length $D$ from the memory replay buffer to train the critic DNN, the target critic DNN, actor DNNs, and target actor DNNs, i.e., update $\theta^{(\text{c})}$ to minimize (\ref{loss_function_critic}), update $\theta^{(\text{c-})}$ with (\ref{soft_update_critic}), update $\theta_{n}^{\text{(a)}}$ with (\ref{actor_n_update_direction}), and update $\theta_{n}^{\text{(a-)}}$ with (\ref{soft_update_actor}). From time slot $T_d+D$, in every $T_u$ time slots, the core network transmits the latest weight vector $\theta_{n}^{\text{(a)}}$ to AP $n$ through the bi-directional backhaul link with $T_d$ time slots delay. AP $n$ receives the latest $\theta_{n}^{\text{(a)}}$ in time slot $2T_d+D+T_u$ and uses it to replace the weight vector $\theta_{n}^{\text{(L)}}$ of local DNN $n$.

In the stage of Repeat ($t> 2T_d+D+T_u$), UE $n$ ($\forall \ n \in \ \mathbb{N}$) observes local state $s_n(t)$ and auxiliary information $o_n(t)$, and transmit them to AP $n$ at the beginning of time slot $t$. AP $n$ sets the transmit power to be  $p_n(t)=\mu_{n}^{\text{(L)}}(s_n(t);\theta_{n}^{\text{(L)}})+\zeta$ and meanwhile uploads local experience $e_n(t)$ and $o_n(t)$ to the core network through the bi-directional backhaul link with $T_d$ time slots delay. After collecting all the local experiences and the corresponding auxiliary information from $N$ APs, the core network constructs a global experience $E(t)$ and stores it into the experience replay buffer. Then, a mini-batch of experiences are sampled from the experience replay buffer to train the critic DNN, the target critic DNN, actor DNNs, and target actor DNNs. In every $T_u$ time slots, AP $n$ receives the latest $\theta_{n}^{\text{(a)}}$ and uses it to replace the weight vector $\theta_{n}^{\text{(L)}}$.

\subsubsection{Discussions on the computational complexity}
The computational complexity at each AP is dominated by the calculation of a DNN with $L_2^{\text{(a)}}+L_3^{\text{(a)}}+9$ neurons, and thus the computational complexity at each AP is around $\mathcal{O}(L_2^{\text{(a)}}+L_3^{\text{(a)}}+9)$. The computational complexity at the core network is dominated by the training of the critic DNN with $N^2+8N+1+L_2^{\text{(S)}}+L_3^{\text{(S)}}+L_{2}^{\text{(A)}}+L_{2}^{\text{(M)}}$ neurons and $N$ actor DNNs. In particular, the critic DNN is first trained and then $N$ actor DNNs can be trained simultaneously. Thus, the computational complexity of each training is around $\mathcal{O}(L_2^{\text{(a)}}+L_3^{\text{(a)}}+10+N^2+8N+L_2^{\text{(S)}}+L_3^{\text{(S)}}+L_{2}^{\text{(A)}}+L_{2}^{\text{(M)}})$. In the simulation, with the designed DNNs, we show that the average time needed to calculate the transmit power at each AP is around $0.34$ ms, which is much less than those by employing WMMSE algorithm and FP algorithm, the average time needed to train the critic DNN is around $9.7$ ms, and the average time needed to train an actor DNN is around $5.9$ ms. It should be pointed that, the computational capability of the nodes in practical networks is much stronger than the computer we use in the simulations. Thus, the average time needed to train a DNN and calculate the transmit power with a DNN can be further reduced in practical networks.

\subsubsection{Discussions on the overheads}
Note that, the proposed architecture contains two kinds of overhead: local information of each AP to the core network and DNN weight vectors from the core network to the AP. The algorithm in [14] needs three kinds of overhead: local information of each AP to the core network, DNN weight vectors from the core network to the AP, and exchanged local information among neighboring APs. This difference means that, the proposed algorithm does not need cooperations among APs for the power control and is much easier than the algorithm in [14] to implement in practical situations. Besides, as the number of APs increases, the number of actor DNNs increases and the resulting overhead may affect the scalability of the proposed architecture. Nevertheless, the increase of APs may also enhance the spectrum utilization efficiency. Therefore, the number of APs needs to be properly designed to balance the overhead and the spectrum utilization efficiency in practical situations.

\subsubsection{Discussions on the Implementations}
It is worth noting that, the proposed algorithm considers three engineering aspects to facilitate the implementation. First, the core network initially does not has any data for learning in practical situations. Then, the proposed algorithm allows each AP to select transmit power randomly to interact with the environment and accumulate useful data. Second, it typically takes some time for the information exchange between APs and the core network in the implementation. Then, the proposed algorithm takes a corresponding transmission latency into considerations and simulation results demonstrate the robustness of the proposed algorithm to this latency. Third, it is demanding for the core network to transmit the weight vectors of actor DNNs to APs for updating the corresponding local DNNs  in each time slot. Then, the proposed algorithm allows the core network to transmit weight vectors periodically and simulation results show that this design does not affect the long-term performance of the proposed algorithm.

%

 \begin{algorithm}[!thp]
\caption{Proposed DRL based multi-agent power control algorithm.}
{\small
\begin{algorithmic}[1]

\STATE \textbf{Initialization:}

\STATE Adopt the structures in Fig. \ref{DNN_structures}-(A) and Fig. \ref{DNN_structures}-(B), establish DNN networks including $\mu_{n}^{\text{(L)}}(s_n;\theta_{n}^{\text{(L)}})$, $\mu_{n}^{\text{(a)}}(s_n;\theta_{n}^{\text{(a)}})$, $\mu_{n}^{\text{(a-)}}(s_n;\theta_{n}^{\text{(a-)}})$, $\mu_{n}^{\text{(L)}}(s_n;\theta_{n}^{\text{(L)}})$, $Q^{\text{(c)}}(s_1,\cdots, s_N,s_{\text{o}}, a_1, \cdots, a_N;\theta^{(\text{c})})$ and $Q^{\text{c-}}(s_1,\cdots, s_N,s_{\text{o}}, a_1, \cdots, a_N;\theta^{(\text{c-})})$.

\STATE Initialize $\theta_{n}^{\text{(L)}}$, $\theta_{n}^{\text{(a)}}$, and $\theta^{(\text{c})}$ randomly, initialize $\theta_{n}^{\text{(a-)}}$ and $\theta^{(\text{c-})}$ with $\theta_{n}^{\text{(a)}}$ and $\theta^{(\text{c})}$, respectively.

\STATE \textbf{Random experience accumulation:}

\STATE At the beginning of time slot $t$ ($t\leq T_d+D$), UE $n$ transmits $s_n(t)$ and $o_n(t)$ to AP $n$, and AP $n$ randomly chooses a transmit power.

\STATE In time slot $t$ ($t\leq T_d+D$), AP $n$ uploads local experience $e_n(t)$ and auxiliary information $o_n(t)$ to the core network. Upon receiving $N$ local experiences together with $N$ auxiliary information, the core network can construct a global experience and store it into the memory replay buffer.

\STATE In time slot $t=T_d+D$, the core network has $D$ global experiences in the memory replay buffer. Then, a mini-batch of experiences are sampled to update $\theta^{(\text{c})}$ to minimize (\ref{loss_function_critic}), update $\theta^{(\text{c-})}$ with (\ref{soft_update_critic}), update $\theta_{n}^{\text{(a)}}$ with (\ref{actor_n_update_direction}), and update $\theta_{n}^{\text{(a-)}}$ with (\ref{soft_update_actor}).

\STATE From time slot $t=T_d+D$, in every $T_u$ time slots, the core network transmits the latest $\theta_{n}^{\text{(a)}}$ to AP $n$. Upon receiving the latest $\theta_{n}^{\text{(a)}}$, AP $n$ use it to replace the weight vector $\theta_{n}^{\text{(L)}}$.

\STATE \textbf{Repeat:}

\STATE At the beginning of time slot $t$ ($t>2T_d+D+T_u$), UE $n$ transmits $s_n(t)$ and $o_n(t)$ to AP $n$, and AP $n$ sets the transmit power to be $p_n(t)=\mu_{n}^{\text{(L)}}(s_n;\theta_{n}^{\text{(L)}})+\zeta$, where $\zeta$ is the action noise.

\STATE In time slot $t$ ($t>2T_d+D+T_u$), AP $n$ uploads local experience $e_n(t)$ and auxiliary information $o_n(t)$ to the core network; a mini-batch of experiences are sampled to update $\theta^{(\text{c})}$ to minimize (\ref{loss_function_critic}), update $\theta^{(\text{c-})}$ with (\ref{soft_update_critic}), update $\theta_{n}^{\text{(a)}}$ ($\forall \ n \in \ \mathbb{N}$) with (\ref{actor_n_update_direction}), and update $\theta_{n}^{\text{(a-)}}$ ($\forall \ n \in \ \mathbb{N}$) with (\ref{soft_update_actor}).

\STATE In every $T_u$ time slots, AP $n$ receives the latest $\theta_{n}^{\text{(a)}}$ and uses it to replaces the weight vector $\theta_{n}^{\text{(L)}}$.
\end{algorithmic}}
\end{algorithm}

\section{Simulation results}

In this section, we provide simulation results to evaluate the performance of the proposed algorithm. For comparison, we have four benchmark algorithms, namely, WMMSE algorithm, FP algorithm, Full power algorithm, Random power algorithm. In particular, the maximum transmit power is used to initialize the WMMSE algorithm and the FP algorithm, which will stop iterations if the difference of the sum-rates per link between two successive iterations is smaller than $0.001$ or the number of iterations is larger than $500$. In the following, we first provide the settings of the simulation and then demonstrate the performance of the propose algorithm as well as the four benchmark algorithms. We implement the proposed algorithm with an open-source software called keras which is based on Tensorflow, and on a computer with the Intel Core i5-8250U and $16$G RAM.

\subsection{Simulation settings}

\begin{table}

\center
\caption{Hyperparameters of each local DNN.}
\footnotesize
\begin{tabular}{|c|c|c|c|c|c|}
\hline
Layers  & $L_1^{\text{(a)}}$ & $L_2^{\text{(a)}}$ & $L_3^{\text{(a)}}$ & $L_4^{\text{(a)}}$ & $L_5^{\text{(a)}}$\\
\hline
Neuron number & $7$ & $100$ & $100$ & $1$ & $1$ \\
\hline
 Activation function & Linear & Relu & Relu & Sigmoid & Linear\\
\hline
Action noise $\zeta$  & \multicolumn{5}{c|}{Normal distribution with zero mean and variance $2$}\cr\cline{2-6}
\hline
\end{tabular}

\center
\caption{Hyperparameters of each actor DNN.}
\footnotesize
\begin{tabular}{|c|c|c|c|c|c|}
\hline
Layers  & $L_1^{\text{(a)}}$ & $L_2^{\text{(a)}}$ & $L_3^{\text{(a)}}$ & $L_4^{\text{(a)}}$ & $L_5^{\text{(a)}}$\\
\hline
Neuron number & $7$ & $100$ & $100$ & $1$ & $1$ \\
\hline
 Activation function & Linear & Relu & Relu & Sigmoid & Linear\\
\hline
 Optimizer  & \multicolumn{5}{c|}{Adam optimizer with learning rate $0.0001$}\cr\cline{2-6}
\hline
Mini-batch size $D$  & \multicolumn{5}{c|}{$128$}\cr\cline{2-6}
\hline
Learning rate $\tau_n^{\text{(a)}}$  & \multicolumn{5}{c|}{$\tau_n^{\text{(a)}}=0.001$}\cr\cline{2-6}
\hline
\end{tabular}

\center
\caption{Hyperparameters of the critic DNN.}
\footnotesize
\begin{tabular}{|c|c|c|c|c|c|c|c|}
\hline
Layers  & $L_1^{\text{(S)}}$ & $L_2^{\text{(S)}}$ & $L_3^{\text{(S)}}$ & $L_1^{\text{(A)}}$ & $L_2^{\text{(A)}}$ & $L_2^{\text{(M)}}$ & $L_3^{\text{(M)}}$\\
\hline
Neuron number & $7N+N^2$ & $200$ & $200$ & $N$ & $200$ & $200$ & $1$ \\
\hline
 Activation function & Linear & Relu & Linear & Linear & Linear & Relu & Linear \\
 \hline
Optimizer  & \multicolumn{7}{c|}{Adam optimizer with learning rate $0.001$}\cr\cline{2-8}
\hline
Mini-batch size $D$  & \multicolumn{7}{c|}{$128$}\cr\cline{2-8}
\hline
Learning rate $\tau^{\text{(c)}} $  & \multicolumn{7}{c|}{$\tau^{\text{(c)}}=0.001$}\cr\cline{2-8}
\hline
Discount factor $\eta$ & \multicolumn{7}{c|}{$0.5$}\cr\cline{2-8}
\hline
\end{tabular}
\vspace{0.1cm}
\end{table}

To begin with, we provide the hyperparameters of the DNNs in Table I, Table II, and Table III, which are determined by cross-validation \cite{HetNet_DL} \cite{HetNet_DRL}. Note that the adopted hyperparameters maybe not the optimal. Since the proposed algorithm with these hyperparameters performs well, we use them to demonstrate the achievable performance rather than the optimal performance of the proposed algorithm. Besides, we pre-process the local/global state information of the proposed algorithm to reduce their variance in the following procedure: we first use the noise power to normalize each channel gain, and then use the mapping function $f(x)=10\log_{10}(1+x)$ to process the data related to the transmit power, channel gain, interference, and SINR.

In the simulations, we consider both two-layer HetNet scenario and three-layer HetNet scenario:
\begin{itemize}
\item Two-layer HetNet scenario: In this scenario, there are five APs whose locations are respectively $(0,0)$, $(500,0)$, $(0,500)$, $(-500, 0)$, and $(0, -500)$ in meters. Each AP has a disc service coverage defined by a minimum distance $\nu_{\text{min}}$ and a maximum distance $\nu_{\text{max}}$ from the AP to the served UE. AP $1$ is in the first layer and AP $n$ $(n\in \ \{2, 3, 4, 5\})$ is in the second layer. $\nu_{\text{min}}$ is set to be $10$ meters for all APs, and $\nu_{\text{max}}$ of AP $1$ in the first layer is $1000$ meters, and $\nu_{\text{max}}$ of each AP in the second layer is $200$ meters. The maximum transmit power of AP $1$ in the first layer is $30$ dBm, the maximum transmit power of each AP in the second layer is $23$ dBm. The served UE by an AP is randomly located within the service coverage of the AP.

\item Three-layer HetNet scenario: In this scenario, there are nine APs whose locations are respectively $(0,0)$, $(500,0)$, $(0,500)$, $(-500, 0)$, $(0, -500)$, $(700, 0)$, $(0, 700)$, $(-700, 0)$, and $(0, -700)$ in meters. AP $1$ is in the first layer, and AP $n$ $(n\in \ \{2, 3, 4, 5\})$ is in the second layer, and AP $n$ $(n\in \ \{6, 7, 8, 9\})$ is in the third layer. $\nu_{\text{min}}$ is set to be $10$ meters for all APs, and $\nu_{\text{max}}$ of AP $1$ in the first layer is $1000$ meters, and $\nu_{\text{max}}$ of each AP in the second layer is $200$ meters, and $\nu_{\text{max}}$ of each AP in the third layer is $100$  meters. The maximum transmit power of AP $1$ in the first layer is $30$ dBm, and the maximum transmit power of each AP in the second layer is $23$ dBm, and the maximum transmit power of each AP in the third layer is $20$ dBm. The served UE by an AP is randomly located within the service coverage of the AP.
\end{itemize}
Furthermore, the transmission bandwidth is set to be $B=10$ MHz, the adopted path-loss model is $120.9 + 37.6 \log10(d)$ in dB, where $d$ in kilometer is the distance between a transmitter and a receiver \cite{Path_loss}, the log-normal shadowing standard deviation is $8$ dB, the noise power $\sigma^2$ at each UE is $-114$ dBm, the delay $T_d$ of the data transmission between the core network and each AP is $T_d=50$ time slots, the period $T_u$ to update the weight vector of each local DNN is $T_u=100$ time slots, and the capacity of the memory replay buffer is $M=1000$.

          \begin{figure}
            \centering
            \includegraphics[scale=0.6]{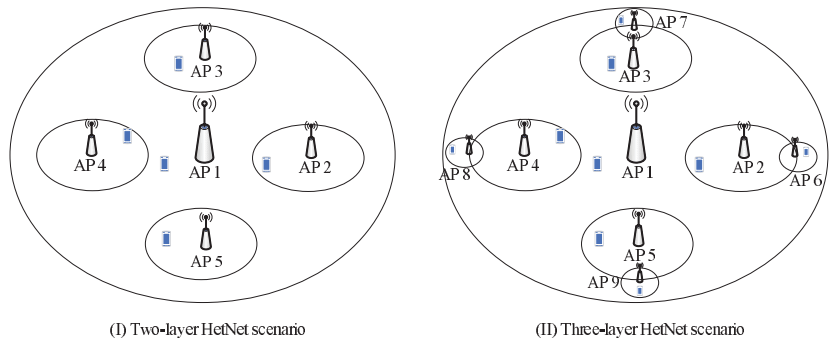}
            \caption{Simulation model: (I) two-layer HetNet scenario; (II) three-layer HetNet scenario.}
             \label{Simulation_model}
        \end{figure}

\subsection{Performance comparison and analysis}
In this part, we will provide the performance comparison and analysis of the proposed algorithm with four benchmark algorithms in two simulation scenarios. In particular, the simulation of the proposed algorithm has two stages, i.e., training stage and testing stage. In the training stage, the DNNs are trained with the proposed Algorithm 1 in the first $5000$ time slots. In the testing stage, the well-trained DNNs are used to optimize the transmit power in each AP in the following $2000$ time slots. Each curve is the average of ten trials, in which the location of the served UE by an AP is randomly generated within the service coverage of the AP.

          \begin{figure}
            \centering
            \includegraphics[scale=1]{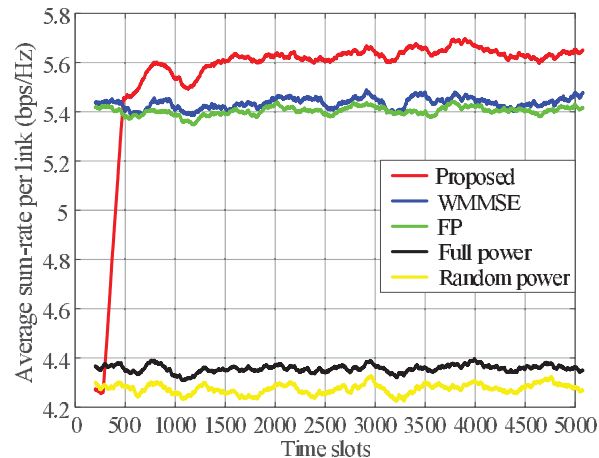}
            \caption{Average sum-rate performance of the two-layer HetNet scenario in the training stage. The channel correlation factor is set to be zero, i.e., IID channel, and each value is the moving average of the previous $200$ time slots.}
            \label{Simulation_two_layer_1000_0}
        \end{figure}

Fig. \ref{Simulation_two_layer_1000_0} provides the average sum-rate performance of the proposed algorithm and four benchmark algorithms in the two-layer HetNet scenario in the training stage. The channel correlation factor is set to be zero, i.e., IID channel. In the figure, the average sum-rate of the proposed algorithm is the same as the random power algorithm at the beginning of data transmissions since the proposed algorithm has to choose transmit power randomly for each AP to accumulate experiences. Then, the average sum-rate of the proposed algorithm increases rapidly and exceeds the sum-rates calculated by WMMSE algorithm and FP algorithm after around $500$ time slots. Finally, the proposed algorithm converges after $1500$ time slots. This can be explained as follows. On the one hand, both WMMSE algorithm and FP algorithm can only output sub-optimal solutions of the power allocation problem in each single time slot, meaning that the average sum-rate performance of both algorithms are also sub-optimal. On the other hand, the proposed algorithm can explore continuously different power control strategies and accumulates global experiences. By learning from these global experiences, the critic DNN has a global view of the impacts of different power control strategies on the sum-rate. Then, the critic DNN can guide each actor DNN (or each local DNN) to update the weight vector towards the global optimum. Thus, it is reasonable that the proposed algorithm outperforms both the WMMSE algorithm and FP algorithm in terms of the average sum-rate.

          \begin{figure}
            \centering
            \includegraphics[scale=1]{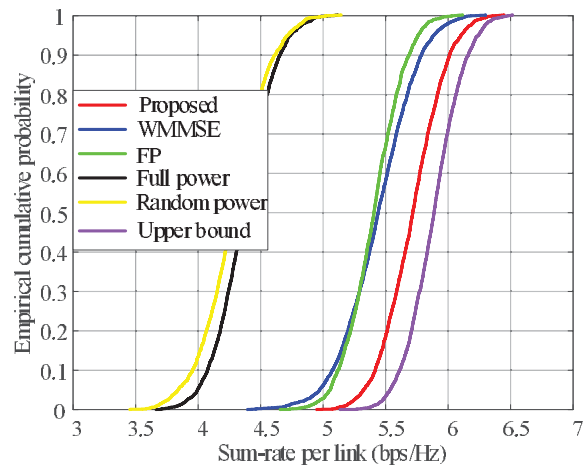}
            \caption{Sum-rate performance of the two-layer intersecting scenario in the testing stage. The channel correlation factor is set to be zero, i.e., IID channel, and each value is the moving average of the previous $200$ time slots.}
            \label{Simulation_two_layer_1000_0_testing}
        \end{figure}

Fig. \ref{Simulation_two_layer_1000_0_testing} provides the corresponding sum-rate performance of the proposed algorithm in the two-layer HetNet in the testing stage. From the figure, the effectiveness of the proposed algorithm in the two-layer HetNet is demonstrated. It should be noted that, the sum-rate performance of the proposed algorithm in the testing stage is also higher that of the proposed algorithm in the training stage. This is because, the proposed algorithm in the training stage needs to continuously explore the transmit power allocation policy and train DNNs until convergence. The exploration may degrade the sum-rate performance after the convergence of the algorithm. On the contrary, by completely exploiting the well-trained DNNs to optimize the transmit power, the proposed algorithm can further enhance the sum-rate performance in the testing stage.


          \begin{figure}
            \centering
            \includegraphics[scale=1]{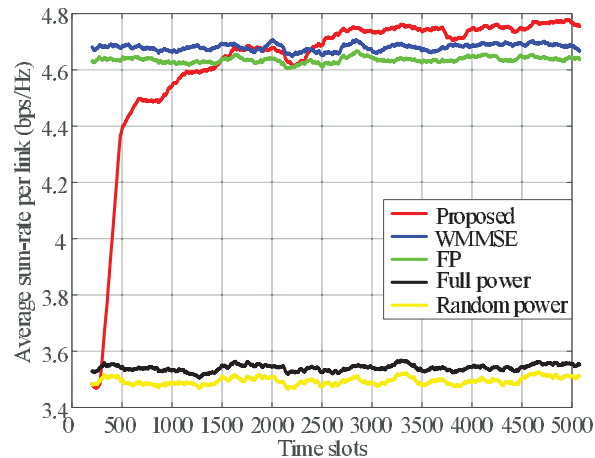}
            \caption{Sum-rate performance of the three-layer HetNet scenario in the training stage. The channel correlation factor is set to be zero, i.e., IID channel, and each value is the moving average of the previous $200$ time slots.}
            \label{Simulation_three_layer_1000_0}
        \end{figure}

          \begin{figure}
            \centering
            \includegraphics[scale=1]{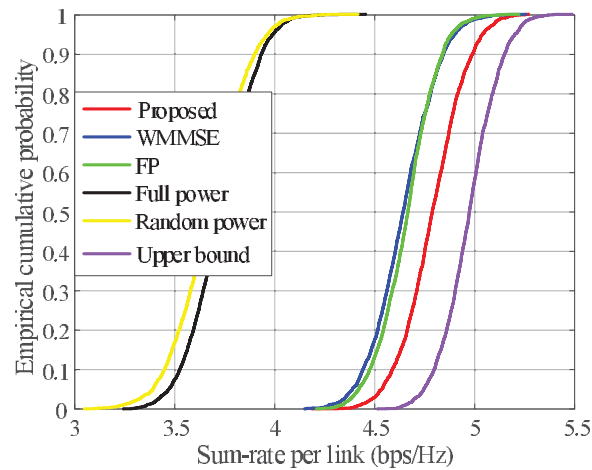}
            \caption{Sum-rate performance in the two-layer overlapping scenario in the testing stage. The channel correlation factor is set to be zero, i.e., IID channel, and each value is the moving average of the previous $200$ time slots.}
            \label{Simulation_three_layer_1000_0_testing}
        \end{figure}

Fig. \ref{Simulation_three_layer_1000_0} provides the average sum-rate performance of the proposed algorithm and four benchmark algorithms in the three-layer HetNet scenario in the training stage. The channel correlation factor is set to be zero, i.e., IID channel. In the figure, the average sum-rate of the proposed algorithm also increases rapidly and exceeds the average sum-rates calculated by WMMSE algorithm and FP algorithm after around $2500$ time slots, and converges after $3000$ time slots. This phenomenon can be explained in a similar way to that in Fig. \ref{Simulation_two_layer_1000_0}. In addition, Fig. \ref{Simulation_three_layer_1000_0_testing} provides the corresponding sum-rate performance of the proposed algorithm in the three-layer HetNet in the testing stage. From the figure, we observe a phenomenon similar to that in Fig. \ref{Simulation_two_layer_1000_0_testing}.


          \begin{figure}
            \centering
            \includegraphics[scale=1]{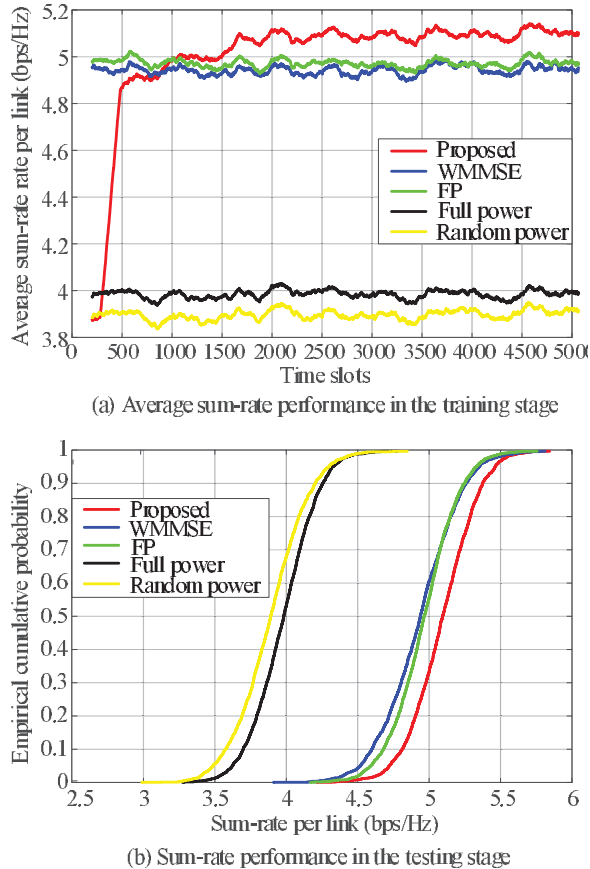}
            \caption{Sum-rate performance of the two-layer HetNet scenario with a random channel correlation factor $\rho$.}
            \label{Simulation_two_layer_1000_random_rho}
        \end{figure}

Fig. \ref{Simulation_two_layer_1000_random_rho} provides the sum-rate performance of the proposed algorithm and four benchmark algorithms in the two-layer HetNet scenario with a random channel correlation factor $\rho$. In Fig. \ref{Simulation_two_layer_1000_random_rho}-(a), the average sum-rate of the proposed algorithm in the training stage increases rapidly and exceeds the average sum-rates calculated by WMMSE algorithm and FP algorithm after around $500$ time slots, and converges after $1000$ time slots. Meanwhile,in Fig. \ref{Simulation_two_layer_1000_random_rho}-(b), the sum-rate of the proposed algorithm in the testing stage is generally higher than those of the benchmark algorithms. This demonstrates that sum-rate performance of the proposed algorithm outperforms those of benchmark algorithms in the two-layer HetNet scenario even with a random $\rho$.


 \begin{figure}
            \centering
            \includegraphics[scale=1]{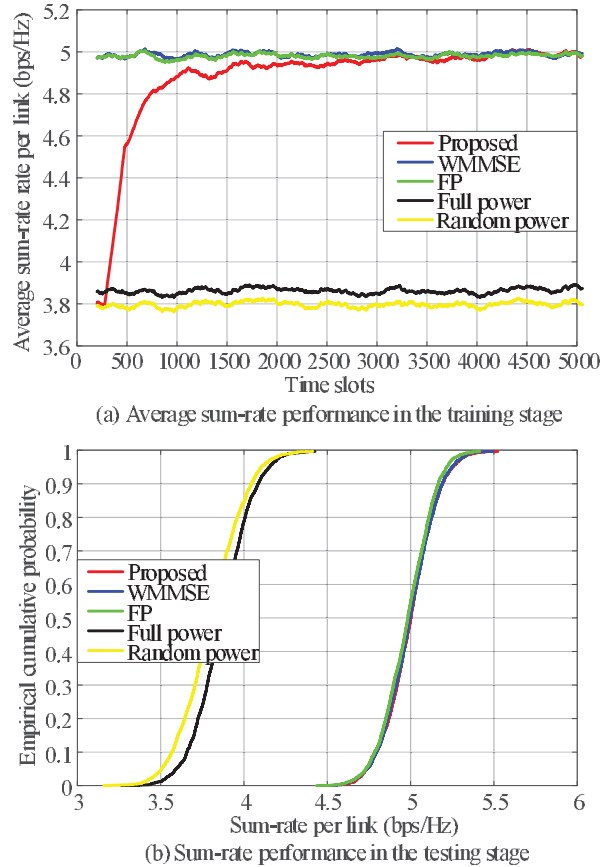}
            \caption{Sum-rate performance in the three-layer HetNet scenario with a random channel correlation factor $\rho$.}
            \label{Simulation_three_layer_1000_random_rho}
        \end{figure}

 Fig. \ref{Simulation_three_layer_1000_random_rho} provides the sum-rate performance of the proposed algorithm and four benchmark algorithms in the three-layer HetNet scenario with a random channel correlation factor $\rho$. In Fig. \ref{Simulation_three_layer_1000_random_rho}-(a), the average sum-rate of the proposed algorithm in the training stage increases rapidly and converges to the average sum-rates calculated by WMMSE algorithm and FP algorithm after around $3000$ time slots. Meanwhile, in Fig. \ref{Simulation_three_layer_1000_random_rho}-(b), the sum-rate of the proposed algorithm in the testing stage is almost the same as those of WMMSE algorithm and FP algorithm. This demonstrates the advantages of the proposed algorithm in the three-layer HetNet scenario even with a random channel correlation factor $\rho$. Note that, the performance advantage of the proposed algorithm in the three-layer HetNet scenario diminishes compared with that in the two-layer HetNet scenario. This is because, the complexity of the network topology increases exponentially as the network size scales up and it is generally more difficult to learn to maximize the sum-rate of the three-layer HetNet scenario than the two-layer HetNet scenario.


\newcommand{\tabincell}[2]{\begin{tabular}{@{}#1@{}}#2\end{tabular}}
 \begin{table}
\center
\caption{Average time complexity.}
\footnotesize
\begin{tabular}{|c|c|c|c|c|}
\hline
\tabincell{c}{Training the\\ critic DNN} & \tabincell{c}{Training an \\ actor DNN} & \tabincell{c}{Calculation with \\ a local DNN} & WMMSE & FP \\
\hline
$9.7$ ms & $5.9$ ms & $0.34$ ms & $120$ ms & $79$ ms \\
\hline
\end{tabular}
\end{table}

Table IV shows the average time complexity of different algorithms in the simulation. From the table, we observe that the average time needed to calculate a transmit power with a local DNN is much less than those of employing WMMSE algorithm and FP algorithm. Besides, the average time needed to train the critic DNN and actor DNNs is below ten mini-seconds. In fact, the computational capability of the nodes in practical networks is much stronger than the computer we use in the simulations. Thus, the average time needed to train a DNN and calculate the transmit power with a DNN will be further reduced in the practical networks.

\section{Conclusions}

In this paper, we exploited DRL to design a multi-agent power control in the HetNet. In particular, a deep neural networks (DNN) was established at each AP and a MASC method was developed to effectively train the DNNs. With the proposed algorithm, each AP can independently learn to optimize the transmit power and enhance the sum-rate with only local information. Simulation results demonstrated the superiority of the proposed algorithm compared with conventional power control algorithms, e.g., WMMSE algorithm and FP algorithm, from the aspects of average sum-rate and computational complexity. In fact, the proposed algorithm framework can also be applied in more resource management problems, in which global instantaneous CSI is unavailable and the cooperations among users are unavailable or highly cost.


\begin{thebibliography}{10}

\bibitem{6GBe} L. Zhang, Y.-C. Liang, and D. Niyato, ``6G visions: Mobile ultra-broadband, super Internet-of-Things, and artificial intelligence,'' \emph{China Communications}, vol. 16, no. 8, pp. 1-14, Aug. 2019.

\bibitem{5GBe} J. G. Andrews, S. Buzzi, W. Choi, S. V. Hanly, A. Lozano, A. C. K. Soong, and J. C. Zhang, ``What will 5G be?'' \emph{IEEE
J. Select. Areas Commun.}, vol. 32, no. 6, pp. 1065-1082, Jun. 2014.

%
\bibitem{Lin_SS_2017} L. Zhang, M. Xiao, G. Wu, M. Alam, Y.-C. Liang, and S. Li, ``A survey of advanced techniques for spectrum sharing in 5G networks,'' \emph{IEEE Wireless Commun.}, vol. 24, no. 5, pp. 44-51, Oct. 2017.
%
%
%
%
\bibitem{HetNet_survey1} M. Agiwal, A. Roy, and N. Saxena, ``Next generation 5G wireless networks: A comprehensive survey,'' \emph{IEEE Commun.
Surv. Tutor.}, vol. 18, no. 3, pp. 1617-1655, Third quarter 2016.
%
%
\bibitem{HetNet_survey2} C. Yang, J. Li, M. Guizani, A. Anpalagan, and M. Elkashlan, ``Advanced spectrum sharing in 5G cognitive heterogeneous networks,'' \emph{IEEE Wireless Commun.}, vol. 23, no. 2, pp. 94-101, Apr. 2016.







\bibitem{HetNet_RA0} S. Singh and J. G. Andrews, ``Joint resource partitioning and offloading in heterogeneous cellular networks,'' \emph{IEEE Wireless Commun.}, vol. 13, no. 2, pp. 888-901, Feb. 2014.

\bibitem{WMMSE} Q. Shi, M. Razaviyayn, Z.-Q. Luo, and C. He, ``An iteratively weighted MMSE approach to distributed sum-utility maximization for a MIMO interfering broadcast channel,'' \emph{IEEE Trans. Signal Process.}, vol. 59, no. 9, pp. 4331-4340, Sep. 2011.

\bibitem{FP} K. Shen and W. Yu, ``Fractional programming for communication systems-Part I: Power control and beamforming,'' \emph{IEEE Trans. Signal Process.}, vol. 66, no. 10, pp. 2616-2630, May 2018.

\bibitem{HetNet_RA3} R. Q. Hu and Y. Qian, ``An energy efficient and spectrum efficient wireless heterogeneous network framework for 5G systems,'' \emph{IEEE Commun. Mag.}, vol. 52, no. 5, pp. 94-101, May 2014.

\bibitem{HetNet_RA4} J. Huang, R. A. Berry, and M. L. Honig, ``Distributed interference compensation for wireless networks,'' \emph{IEEE J. Sel. Areas Commun.}, vol. 24, no. 5, pp. 1074-1084, May 2006.

\bibitem{HetNet_RA4} H. Zhang, L. Venturino, N. Prasad, P. Li, S. Rangarajan, and X. Wang, ``Weighted sum-rate maximization in multi-cell networks via coordinated scheduling and discrete power control,'' \emph{IEEE J. Sel. Areas Commun.}, vol. 29, no. 6, pp. 1214-1224, Jun. 2011.

\bibitem{HetNet_game} L. B. Le and E. Hossain, ``Resource allocation for spectrum underlay in cognitive radio networks,'' \emph{IEEE Trans. Wireless Commun.}, vol. 7, no. 12, pp. 5306-5315, Dec. 2008.



\bibitem{HetNet_DL} H. Sun, X. Chen, Q. Shi, M. Hong, X. Fu, and N. D. Sidiropoulos, ``Learning to optimize: Training deep neural networks for interference management,'' \emph{IEEE Trans. Signal Process.}, vol. 66, no. 20, pp. 5438-5453, Oct. 2018.

\bibitem{HetNet_DRL} Y. S. Nasir and D. Guo, ``Multi-agent deep reinforcement learning for dynamic power allocation in wireless networks,'' \emph{IEEE J. Sel. Areas in Commun.}, vol. 37, no. 10, pp. 2239-2250, Oct. 2019.

\bibitem{DRL_0} L. Xiao, H. Zhang, Y. Xiao, X. Wan, S. Liu, L.-C. Wang, and H. V. Poor, ``Reinforcement learning-based downlink interference control for ultra-dense small cells,'' \emph{IEEE Trans. Wireless Commun.}, vol. 19, no. 1, pp. 423-434, Jan. 2020.

\bibitem{DRL_1} R. Amiri, M. A. Almasi, J. G. Andrews, and H. Mehrpouyan, ``Reinforcement learning for self organization and power Control of two-tier heterogeneous networks,'' \emph{IEEE Trans. Wireless Commun.}, vol. 18, no. 8, pp. 3933-3947, Aug. 2019.



\bibitem{RL_0} Y. Sun, G. Feng, S. Qin, Y.-C. Liang, and T.-S. P. Yum, ``The SMART handoff policy for millimeter wave heterogeneous cellular networks,'' \emph{IEEE Trans. Mobile Commun.}, vol. 17, no. 6, pp. 1456-1468, Jun. 2018.

\bibitem{RL_1} D. D. Nguyen, H. X. Nguyen, and L. B. White, ``Reinforcement learning with network-assisted feedback for heterogeneous RAT selection,'' \emph{IEEE Trans. Wireless Commun.}, vol. 16, no. 9, pp. 6062-6076, Sep. 2017.

\bibitem{RL_2} Y. Wei, F. R. Yu, M. Song, and Z. Han, ``User scheduling and resource allocation in HetNets with hybrid energy supply: an actor-critic reinforcement learning approach,'' \emph{IEEE Trans. Wireless Commun.}, vol. 17, no. 1, pp. 680-692, Jan. 2018.

\bibitem{RL_3} N. Morozs, T. Clarke, and D. Grace, ``Heuristically accelerated reinforcement learning for dynamic secondary spectrum sharing,'' \emph{IEEE Access}, vol. 3, pp. 2771-2783, 2015.


\bibitem{RL_4} V. Raj, I. Dias. T. Tholeti, and S. Kalyani, ``Spectrum access in cognitive radio using a two-stage reinforcement learning approach,'' \emph{IEEE J. Sel. Topics Signal Process.}, vol. 12, no. 1, pp. 20-34, Feb. 2018.

\bibitem{RL_5} O. Iacoboaiea, B. Sayrac, S. B. Jemaa, and P. Bianchi, ``SON coordination in heterogeneous networks: a reinforcement learning framework,'' \emph{IEEE Trans. Wireless Commun.}, vol. 15, no. 9, pp. 5835-5847, Sep. 2016.

%
%

\bibitem{DRL_nature} V. Mnih et al., ``Human-level control through deep reinforcement learning,'' \emph{Nature}, vol. 518, no. 7540, pp. 529-533, 2015.

\bibitem{DL_survey} C. Zhang, P. Patras, and H. Haddadi,``Deep Learning in Mobile and Wireless Networking: A Survey,'' \emph{IEEE Commun. Surveys Tuts.}, vol. 21, no. 3, pp. 2224-2287, 3rd Quart., 2019.

\bibitem{DL_PC} T. V. Chien, T. N. Canh, E. Bjornson, and E. G. Larsson, ``Power control in cellular massive MIMO with varying user
activity: A deep learning solution,'' \emph{IEEE Trans. Wireless Commun.}, DOI: 10.1109/TWC.2020.2996368.

\bibitem{DL_CA} R. Mennes, F. A. P. D. Figueiredo, and S. Latre, ``Multi-agent deep learning for multi-channel access in slotted wireless networks,'' IEEE Access, vol. 8, pp. 95032-95045, 2020.

\bibitem{DL_LS} W. Cui, K. Shen, and W. Yu, ``Spatial deep learning for wireless scheduling,'' \emph{IEEE J. Select. Areas Commun.}, vol. 37, no. 6, pp. 1248-1261, Jun. 2019.


\bibitem{DRL_2} H. Ye, G. Y. Li, and B.-H. F. Juang, ``Deep reinforcement learning based resource allocation for V2V communications,'' \emph{IEEE Trans. Veh. Technol.}, vol. 68, no. 4, pp. 3163-3173, Apr. 2019.

\bibitem{DRL_3} Y. Yu, T. Wang, and S. C. Liew, ``Deep-reinforcement learning multiple access for heterogeneous wireless networks,'' \emph{IEEE J. Sel. Areas Commun.}, vol. 37, no. 6, pp. 1277-1290, Jun. 2019.

\bibitem{DRL_4} Y. He, Z. Zhang, F. R. Yu, N. Zhao, H. Yin, V. C. M. Leung, and Y. Zhang, ``Deep-reinforcement-learning-based optimization for cache-enabled opportunistic interference alignment wireless networks,'' \emph{IEEE Trans. Veh. Technol.}, vol. 66, no. 11, pp. 10433-10445, Sep. 2017.

\bibitem{DRL_5} L. Zhang, J. Tan, Y.-C. Liang, G. Feng, and D. Niyato, ``Deep reinforcement learning-Based modulation and coding scheme selection in cognitive heterogeneous networks,'' \emph{IEEE Trans. Wireless Commun.}, vol. 18, no. 6, pp. 3281-3294, Jun. 2019.

\bibitem{DRL_6} F. B. Mismar, B. L. Evans, and A. Alkhateeb, ``Deep reinforcement learning for 5G networks: Joint beamforming, power control, and interference coordination,'' \emph{IEEE Trans. Commun.}, vol. 68, no. 3, pp. 1581-1592, Mar. 2020.

\bibitem{DRL_7} H. Huang, Y. Yang, H. Wang, Z. Ding, H. Sari, and F. Adachi, ``Deep reinforcement learning for UAV navigation through massive MIMO technique,'' \emph{IEEE Trans. Veh. Technol.}, vol. 69, no. 1, pp. 1117-1121, Jan. 2020.

\bibitem{DRL_8} H. Zhang, N. Yang, W. Huangfu, K. Long, and V. C. M. Leung, ``Power control based on deep reinforcement learning for spectrum sharing,'' \emph{IEEE Trans. Wireless Commun.}, vol. 19, no. 6, pp. 4209-4219, Jun. 2020.


\bibitem{DRL_survey} N. C. Luong et al., ``Applications of deep reinforcement learning in communications and networking: A survey,'' IEEE Commun. Surveys, vol. 21, no. 4, pp. 3133-3174, 2019.


\bibitem{Jake_model} T. Kim, D. J. Love, B. Clerckx, ``Does frequent low resolution feedback outperform infrequent high resolution feedback for multiple antenna beamforming systems?'', \emph{IEEE Trans. Signal Process.}, vol. 59, no. 4, pp. 1654-1669, Apr. 2011.

\bibitem{NP_Hard} Z.-Q. Luo and S. Zhang, ``Dynamic spectrum management: Complexity and duality,'' \emph{IEEE J. Sel. Topics Signal Process.}, vol. 2, no. 1, pp. 57-73, Feb. 2008.

\bibitem{DDPG} T. P. Lillicrap, J. J. Hunt, A. Pritzel, N. Heess, T. Erez, Y. Tassa, D. Silver, and D. Wierstra, ``Deterministic Policy Gradient Algorithms,'' ICML, Jun. 2016.

\bibitem{CCDRL} T. P. Lillicrap, J. J. Hunt, A. Pritzel, N. Heess, T. Erez, Y. Tassa, D. Silver, and D. Wierstra, ``Continuous control with deep reinforcement learning,'' ICLR, Aug. 2015.



\bibitem{Path_loss} Radio Frequency (RF) System Scenarios, document 3GPP TR 25.942, v.14.0.0, 2017.


%
%
%
%
%
%
%
%
%


%
%
%
%
%
%

%
%
%
%
%
%
%
%
%


\end{thebibliography}
\end{document}